\documentclass[%
 reprint,
 amsmath,amssymb,
 aps,
 prd,
]{revtex4-2}

\usepackage{graphicx}% Include figure files
\usepackage{dcolumn}% Align table columns on decimal point
\usepackage{bm}% bold math
\usepackage{appendix}
\usepackage{makecell}

\begin{document}

\title{Prospects for Distinguishing Supernova Models Using a Future Neutrino Signal}
\author{Jackson Olsen}
\author{Yong-Zhong Qian}
\affiliation{School of Physics and Astronomy, University of Minnesota, Minneapolis, Minnesota 55455}

\date{\today}% It is always \today, today,
             %  but any date may be explicitly specified

\begin{abstract}
The next Galactic core-collapse supernova (SN) should yield a large number of observed neutrinos. Using Bayesian techniques, 
we show that with an SN at a known distance up to 25~kpc, the neutrino events in a water Cherenkov detector similar to Super-Kamiokande 
(SK) could be used to distinguish between seven one-dimensional neutrino emission models assuming no flavor oscillations or the standard 
Mikheyev-Smirnov-Wolfenstein effect. Some of these models could still be differentiated with an SN at a known distance of 50~kpc. 
We also consider just the relative distributions of neutrino energy and arrival time predicted by 
the models and find that a detector like SK meets the requirement to distinguish between these distributions with an SN at an unknown 
distance up to $\sim 10$~kpc.
\end{abstract}

\maketitle

\section{\label{intro}Introduction}
Since the observation of Supernova 1987A (SN 1987A) and its associated neutrino signal, 
theoretical work on core-collapse SNe and their neutrino emission has advanced significantly
(see, e.g., \cite{janka2012,Mirizzi:2015eza} for recent reviews). 
Today, both one-dimensional (1D) and multi-D (2D and 3D) models of SN neutrino emission can be simulated
beyond $\sim 1$~s (e.g., \cite{Mirizzi:2015eza,nagakura}).
In \cite{Olsen2021}, we took the Bayesian approach to compare three 1D models provided by 
the Garching group \cite{Garching} with the SN 1987A data from the Kamiokande II (KII) detector 
\cite{PhysRevLett.58.1490,PhysRevD.38.448}. While providing some discrimination among
the models, the sparse KII data prevented us from drawing definitive conclusions.
More useful will be a future neutrino signal from a Galactic SN, which is expected to 
result in a great many neutrino events in current and planned detectors (e.g., \cite{Scholberg}). 
The actual number of events will depend on the distance to the SN, the details of the neutrino emission,
and the detector of concern. 

The authors of \cite{Hyper-Kamiokande:2021frf} performed a detailed study of the feasibility of 
distinguishing between five SN models using hypothetical neutrino data in the planned Hyper-Kamiokande 
(HK) detector (e.g., \cite{Hyper-Kamiokande:2018ofw}). They focused on an early phase of neutrino emission
spanning 500 ms, which is closely related to the explosion mechanism. Assuming that the distance to the
SN is unknown, they studied the relative distributions of neutrino energy and arrival time predicted by 
the models. Taking into account flavor oscillations due to the Mikheyev-Smirnov-Wolfenstein (MSW) effect,
they performed detailed reconstruction of 100 and 300 simulated events
in the HK detector.
In addition to the dominant inverse beta decay (IBD) reaction of $\bar\nu_e$ on protons, they also 
included scattering of all neutrino species on electrons and charged-current reactions of $\nu_e$ and
$\bar\nu_e$ on $^{16}$O. They concluded that all the five models could be distinguished with 300 events, 
which are expected from an SN at a distance of $\sim 60$--100 kpc for these models. A further analysis
similar to the above was carried out in \cite{Migenda:2019xbm} to study the feasibility of distinguishing
between four models that differ specifically in the mass or initial metallicity of the SN progenitor.

In this paper, we present a complementary study to those of \cite{Hyper-Kamiokande:2021frf,Migenda:2019xbm}.
We focus on seven 1D SN models provided by the Garching group \cite{Garching}. None of these models were
considered in \cite{Hyper-Kamiokande:2021frf,Migenda:2019xbm}. Each model covers 
neutrino emission for at least $9$~s, which allows us to explore the proto-neutron star (PNS) cooling phase 
in addition to the accretion phase of interest to \cite{Hyper-Kamiokande:2021frf,Migenda:2019xbm}.
To cover the range of possible characteristics of water Cherenkov detectors, we consider both 
a detector similar to Super-Kamiokande (SK) and an idealized version. Following the treatments used extensively 
in the analyses of the SN 1987A neutrino data (e.g., \cite{Olsen2021, Loredo:2001rx}), we consider only 
the dominant IBD detection channel. We allow for the Poisson statistics of the number of neutrino events 
and study the discriminating power of our assumed detectors for an SN at a specific distance. We analyze 
separately the cases where the distance to the SN is known or unknown.
In addition to the MSW effect with the normal or inverted neutrino mass hierarchy
considered in \cite{Hyper-Kamiokande:2021frf,Migenda:2019xbm}, we also include the case of no oscillations 
for reference. We employ the Bayesian statistics to test the distinguishability of pairs of SN models
for each of these three oscillation cases, and study the feasibility of distinguishing between 
these cases for a specific SN model. We find that all seven models can be distinguished from each other by
an SK-like detector with an SN at a known distance up to at least 25~kpc or at an unknown distance up to at least $\sim 10$~kpc.
In addition, provided that the underlying model is known, the three oscillation cases can be distinguished 
from each other with our assumed ideal detector and an SN at a known distance of 10~kpc.

The rest of the paper proceeds as follows. In Sec.~\ref{models} we describe the seven SN neutrino emission models 
and the two hypothetical detectors used in this work. In Sec.~\ref{example} we first perform a Monte Carlo study of 
the signal from one SN model and illustrate our general methodology of using the Bayes factor to distinguish a pair 
of models. We then present the mean Bayes factors and the associated standard deviations for various pairs of models
assuming an SN at several known distances. In Sec.~\ref{prefactor} we calculate the combination of detector mass, 
detection efficiency, and SN distance required to distinguish between each pair of models assuming an unknown distance
to the SN. We summarize our results and give conclusions in Sec.~\ref{concl}. 

\section{\label{models}Neutrino Emission and Detection}
We employ seven models of SN neutrino emission in our analysis, all of which are 1D simulations provided 
by the Garching group \cite{Garching}. They differ in the progenitor mass and the nuclear equation of state 
(EoS) used, and are designated as z9.6-LS220, z9.6-SFHo, s18.6-LS220, s18.6-SFHo, s20-SFHo, s27-LS220, and s27-SFHo.
The z9.6, s18.6, s20, and s27 models correspond to progenitor masses of $9.6$, $18.6$, $20$, and $27\,M_\odot$, 
respectively. The designation LS220 or SFHo corresponds to the EoS of \cite{LSEoS} or \cite{SFHoEoS}, respectively. 
To keep our analysis consistent across the models, we use $9$~s of neutrino emission. Some of these models were 
described in detail in \cite{Mirizzi:2015eza}.

The progenitor mass mainly influences the accretion phase of neutrino emission, during which matter falls onto 
the PNS before shock revival, releasing primarily $\nu_e$ and $\bar\nu_e$. The density of the infalling matter
depends on the progenitor structure. The slower decrease of density with radius for more massive progenitors 
delays the shock revival to later times, and therefore, leads to a longer accretion phase (up to $\sim 0.6$~s
for s20 and s27 models). The EoS mainly influences the emission due to cooling of the PNS, which
lasts $\gtrsim 10$~s. For convenience, we refer to the period after the accretion phase as the PNS cooling
phase, although PNS cooling starts at the same time as accretion-induced emission. In contrast to the
accretion phase with dominant emission of $\nu_e$ and $\bar\nu_e$, the PNS cooling phase
is characterized by approximately equal luminosities for $\nu_e$, $\bar\nu_e$, $\nu_x$, and $\bar\nu_x$ 
($x=\mu$ or $\tau$). 

\begin{figure*}
\includegraphics[width=2\columnwidth]{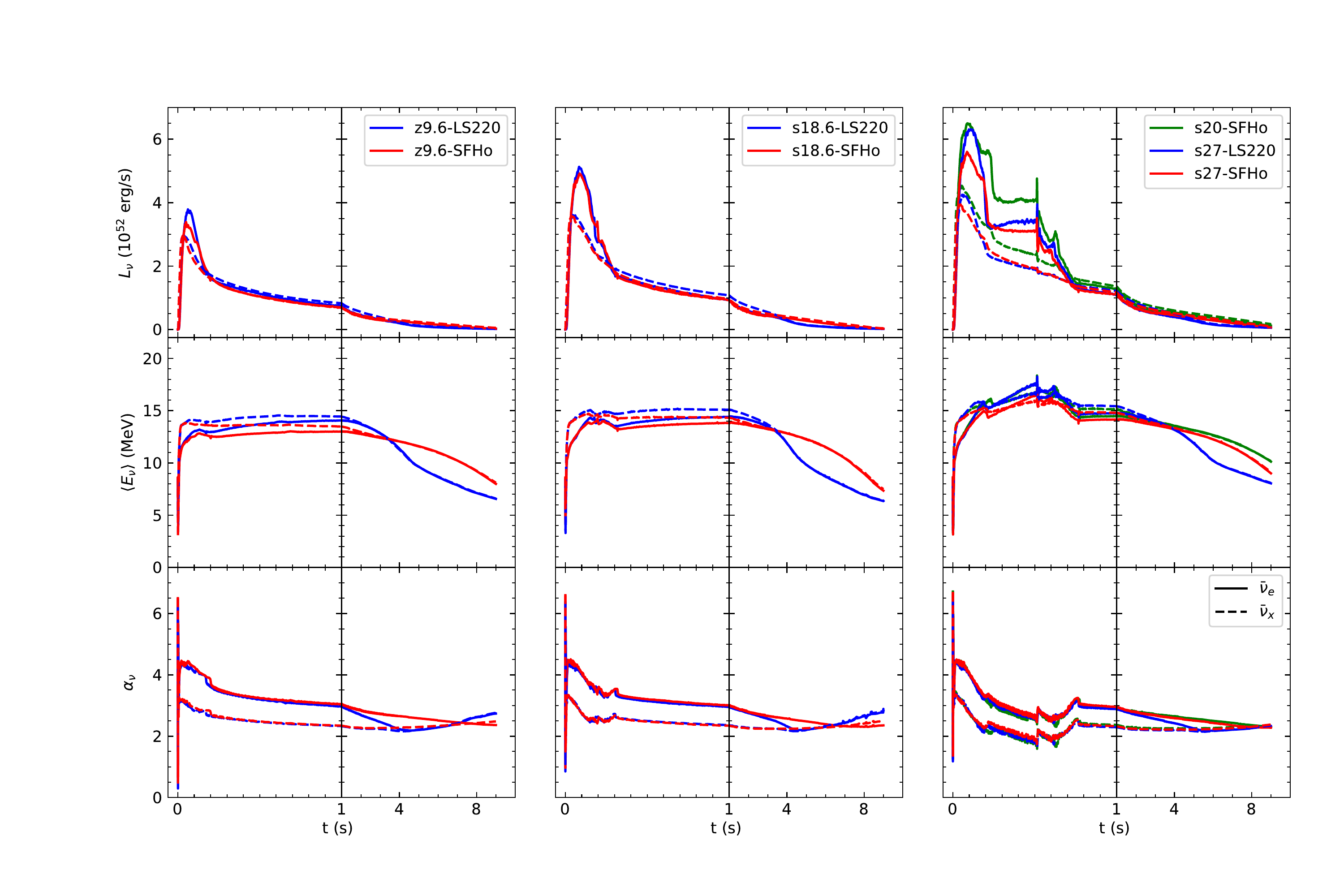}
\caption{The luminosity $L_\nu$, average energy $\langle E_\nu \rangle$, and spectral parameter $\alpha_\nu$ for $\bar\nu_e$ and $\bar\nu_x$ are displayed as functions of time for the adopted SN neutrino emission models. Note that the time scale changes at $t=1$~s.}
\label{fig:models}
\end{figure*}

Figure~\ref{fig:models} shows the characteristics of $\bar\nu_e$ and $\bar\nu_x$ emission as functions of time
for our adopted models. The evolution of $\bar\nu_e$ and $\bar\nu_x$ luminosities, 
$L_{\bar\nu_e}$ and $L_{\bar\nu_x}$, is shown in the first row. 
The z9.6 models, with their low progenitor mass, have a very short accretion phase with only a little excess 
emission of $\bar\nu_e$ over $\bar\nu_x$. Differences between $L_{\bar\nu_e}$ and $L_{\bar\nu_x}$ for 
the s18.6 models indicate an accretion phase of moderate duration and intensity, while the pronounced 
differences for the s20 and s27 models reveal long and intense accretion-induced emission of $\bar\nu_e$. 
For all seven models, we see only small differences between $L_{\bar\nu_e}$ and $L_{\bar\nu_x}$
during the PNS cooling phase.
The second row of Fig.~\ref{fig:models} shows the evolution of the average $\bar\nu_e$ and $\bar\nu_x$ energies,
$\langle E_{\bar\nu_e}\rangle$ and $\langle E_{\bar\nu_x}\rangle$.
We see that $\langle E_{\bar\nu_x}\rangle$ is larger than $\langle E_{\bar\nu_e}\rangle$ initially, 
but the difference subsides after several seconds. We also see that the LS220 and SFHo models differ in
that the former have a quicker drop of $\langle E_{\bar\nu_e}\rangle$ and $\langle E_{\bar\nu_x}\rangle$
during the PNS cooling phase.
The third row of Fig.~\ref{fig:models} shows the evolution of the spectral parameter
\begin{equation}
\alpha_{\nu}=\frac{2 \langle E_{\nu} \rangle^2 - \langle E_{\nu}^2 \rangle}
{\langle E_{\nu}^2 \rangle - \langle E_{\nu} \rangle^2}
\end{equation}
for $\bar\nu_e$ and $\bar\nu_x$, where $\langle E_{\nu}^2 \rangle$ is the second moment of the neutrino 
energy spectrum. We see that $\alpha_{\bar\nu_e}$ and $\alpha_{\bar\nu_x}$ differ significantly 
at early times, but grow more similar at late times. The difference between $\alpha_{\bar\nu_e}$ and 
$\alpha_{\bar\nu_x}$ is insensitive to the EoS at early times, but shows some dependence on the EoS
at late times.

In the absence of neutrino oscillations, the energy-differential number flux of a neutrino species 
$\nu_\beta$ at a distance $d$ to the SN is
\begin{equation}
    F_{\nu_\beta}(E_\nu,t) = \frac{L_{\nu_\beta}}{4 \pi d^2\langle E_{\nu_\beta}\rangle}f_{\nu_\beta}(E_\nu,t),
\end{equation}
where
\begin{equation}
    f_{\nu_\beta}(E_\nu,t)=\frac{T_{\nu_\beta}^{-1}(t)}{\Gamma(1 + \alpha_{\nu_\beta}(t))} 
    \left(\frac{E_\nu}{T_{\nu_\beta}(t)}\right)^{\alpha_{\nu_\beta}(t)} e^{-E_\nu/T_{\nu_\beta}(t)},
\end{equation}
with $\Gamma(1+\alpha_{\nu_\beta})$ being the Gamma function and
\begin{equation}
    T_{\nu_\beta}=\frac{\langle E_{\nu_\beta}\rangle}{1 + \alpha_{\nu_\beta}},
\end{equation}
is the normalized $\nu_\beta$ energy spectrum \cite{Tamborra2012}. 
Note that although $L_{\nu_\beta}$,
$\langle E_{\nu_\beta}\rangle$, $\alpha_{\nu_\beta}$, and $T_{\nu_\beta}$ are functions of time, we often suppress their time dependence for convenience.
Here and below, we use the emission
time as the effective arrival time because the time of travel over a fixed distance only introduces
a constant shift.

Because the IBD detection channel, $\bar\nu_e+p\to n+e^+$, has a cross section much larger 
than that of any other channel at the relevant neutrino energies 
\cite{Scholberg,PhysRevD.36.2283,STRUMIA200342}, it will yield the highest number of 
SN neutrino events in water Cherenkov detectors. Most, if not all, of the neutrino events from SN 1987A 
were observed via this channel in such detectors \cite{Scholberg,PhysRevD.38.448,PhysRevLett.58.1494}.
With the recent addition of gadolinium in the detector, SK should have the capability of tagging the IBD 
events from a future SN \cite{Super-Kamiokande:2021the}. Based on the above, we restrict our analysis 
in this paper to the IBD events in water Cherenkov detectors.

The expected energy-differential rate of IBD events including both the signal and the background is
\begin{equation}
\begin{split}
    \frac{d^2N}{dtdE}(E,t) = B(E) + N_p\int F_{\rm det}(E_\nu,t)\sigma_{\rm IBD}(E_\nu)\\
    \times\frac{\epsilon(E_e)}{\sigma_E\sqrt{2 \pi}}
    \exp\left[-\frac{(E-E_e)^2}{2 \sigma_E^2}\right]dE_\nu,
\end{split}
\label{eq:eventrate}
\end{equation}
where $B(E)$ is the background rate at energy $E$, $N_p$ is the total number of free protons within the
fiducial volume, $F_{\rm det}(E_\nu,t)$ is the $\bar\nu_e$ flux at the detector, $\sigma_{\rm IBD}(E_\nu)$ 
is the IBD cross section, $E_e=E_\nu-\Delta$ is the energy of the $e^+$ from the IBD reaction, 
$\Delta=1.293$~MeV is the neutron-proton mass difference, and $\epsilon(E_e)$ is the detection efficiency. 
Because of smearing, an $e^+$ of energy $E_e$ may be detected at energy $E$, the probability of which is
approximated by a Gaussian distribution with an $E_e$-dependent standard deviation $\sigma_E$.

The flux $F_{\rm det}(E_\nu,t)$ is affected by neutrino oscillations and is given by
\begin{equation}
    F_{\rm det}(E_\nu,t) = f F_{\bar\nu_e}(E_\nu,t) + (1-f) F_{\bar\nu_x}(E_\nu,t),
\end{equation}
where the constant $f$ specifies the degree of mixing between $\bar\nu_e$ and $\bar\nu_x$. 
We consider three cases of neutrino oscillations. The reference case NO with no oscillations corresponds 
to $f=1$. The other two cases correspond to $f=0.681$ or $0.022$ for just the MSW effect with the normal
(NH) or inverted (IH) neutrino mass hierarchy, respectively \cite{PhysRevD.62.033007,Gonzalez-Garcia,Zyla:2020zbs}. 
We add (NO), (NH), or (IH) to the label of a model to specify the assumed case of neutrino oscillations.

For a specific emission model $M_\alpha$, the probability distribution for an event to be observed 
at time $t$ with energy $E$ is
\begin{equation}
    p(E, t|M_\alpha) = \frac{1}{\langle N \rangle}\frac{d^2N}{dtdE},
    \label{eq:etdistr}
\end{equation}
where
\begin{equation}
    \langle N \rangle =\int_0^{\rm 9~s} dt\int_{E_{\rm min}}^{\infty} dE \frac{d^2N}{dtdE}
\end{equation}
is the expected total number of events and $E_{\rm min}$ is the minimum energy for detection.

In the above discussion of SN neutrino detection, the quantities $B(E)$, $N_p$, $\epsilon(E_e)$, $\sigma_E$, 
and $E_{\rm min}$ depend on the detector. We consider two hypothetical detectors to cover a range of
capabilities. One detector has a constant detection efficiency $\epsilon=0.75$ and no background [$B(E)=0$]
when a minimum detected energy $E_{\min}=7.5$~MeV is imposed. Its energy resolution is specified by
the standard deviation $\sigma_E$ for the smearing of the $e^+$ energy as
\begin{equation}
\frac{\sigma_E}{\rm MeV}=-0.0839 + 0.349\sqrt{\frac{E_e}{\rm MeV}} + 0.0397\left(\frac{E_e}{\rm MeV}\right).
\end{equation}
Because these characteristics are similar to those of SK (e.g., \cite{Super-K-IV}), we refer to the
above detector as the SK-like detector. We also consider an ideal detector that has no background [$B(E)=0$]
and can detect any $e^+$ above the threshold for Cherenkov radiation ($E_{\rm min} = 0.8$~MeV) with perfect detection
efficiency ($\epsilon=1$) and energy resolution [$\sigma_E=0$, for which the Gaussian distribution becomes
$\delta(E-E_e)$]. 

For both our hypothetical detectors, the event rate in Eq.~(\ref{eq:eventrate}) can be rewritten as
\begin{equation}
\begin{split}
    \frac{d^2N}{dtdE}(E,t) = A\int {\tilde F}_{\rm det}(E_\nu,t)\sigma_{\rm IBD}(E_\nu)\\
    \times\frac{1}{\sigma_E\sqrt{2 \pi}}
    \exp\left[-\frac{(E-E_e)^2}{2 \sigma_E^2}\right]dE_\nu,
\end{split}
\label{eq:eventratea}
\end{equation}
where ${\tilde F}_{\rm det}(E_\nu,t)=F_{\rm det}(E_\nu,t)(d/{\rm kpc})^2$ is the detected flux 
from an SN at a distance of 1~kpc,
\begin{equation}
    A=\epsilon N_p\left(\frac{\rm kpc}{d}\right)^2
    =2.14\times 10^{33}\epsilon\left(\frac{M_{{\rm H}_2{\rm O}}}{32\ {\rm kton}}\right)
    \left(\frac{\rm kpc}{d}\right)^2,
\label{eq:afact}
\end{equation}
and $M_{{\rm H}_2{\rm O}}$ is the fiducial mass of water in the detector. We take
$M_{{\rm H}_2{\rm O}}=32$~kton (appropriate for SK \cite{Scholberg}) for our calculations. 
The expected total numbers of events $\langle N\rangle$ in our hypothetical detectors from
an SN at a distance of 10~kpc are given in Table~\ref{tab:predicted_event_numbers} for
our adopted SN models and assumed cases of neutrino oscillations.
We note that here and below, our results also apply to other combinations of 
$\epsilon$, $M_{{\rm H}_2{\rm O}}$, and $d$ so long as they give the same values of $A$ 
corresponding to our results.

\begin{table}
    \caption{Expected numbers of IBD events $\langle N \rangle$ in our SK-like and ideal detectors 
    from an SN at a distance of 10~kpc for adopted SN models and assumed cases of neutrino 
    oscillations.}
    \label{tab:predicted_event_numbers}
    \centering
    \begin{tabular}{ccc}
    \hline
    \hline
    Model & $\langle N \rangle$ (SK-like) & $\langle N \rangle$ (Ideal) \\
    \hline
    z9.6-LS220&&\\
    (NO) & 2502.54 & 3553.15 \\
    (NH) & 2651.36 & 3755.53 \\
    (IH) & 2958.80 & 4173.62 \\
    z9.6-SFHo&&\\
    (NO) & 2522.06 & 3611.22 \\
    (NH) & 2645.24 & 3777.23 \\
    (IH) & 2899.72 & 4120.17 \\
    s18.6-LS220&&\\
    (NO) & 3517.01 & 4956.11 \\
    (NH) & 3716.93 & 5222.31 \\
    (IH) & 4129.93 & 5772.23 \\
    s18.6-SFHo&&\\
    (NO) & 3763.67 & 5312.96 \\
    (NH) & 3911.38 & 5511.05 \\
    (IH) & 4216.54 & 5920.27 \\
    s20-SFHo&&\\
    (NO) & 7152.88 & 9938.71 \\
    (NH) & 7098.49 & 9863.93 \\
    (IH) & 6986.13 & 9709.45 \\
    s27-LS220&&\\
    (NO) & 5529.75 & 7693.13 \\
    (NH) & 5503.49 & 7655.92 \\
    (IH) & 5449.24 & 7579.06 \\
    s27-SFHo&&\\
    (NO) & 5574.73 & 7786.92 \\
    (NH) & 5608.19 & 7829.64 \\
    (IH) & 5677.30 & 7917.90 \\
    \hline
    \hline
    \end{tabular}
\end{table}

\section{\label{example}Analysis for Known SN Distance}
We now present a Bayesian approach to test the distinguishability 
of our adopted SN models with an SN at a known distance.
We first perform a Monte Carlo study of the signal from one model
and illustrate our general methodology of using the Bayes factor
to distinguish a pair of models. We then present the mean
Bayes factors and the associated standard deviations for various pairs of
models assuming an SN at several known distances.

\subsection{\label{montecarlo}An Example}
As a demonstration of our Bayesian approach, we consider the following example.
We assume that an SN occurs at $d=50$~kpc (approximately the distance of SN 1987A 
\cite{1991ApJ...380L..23P,Panagia:2003rt}) with its neutrino emission described 
by the model z9.6-LS220(NO). To test how well we can distinguish between this 
true model and any other model, we generate a Monte Carlo sample of $10^4$
instances of the signal in our assumed ideal detector from the above SN. For each 
simulated signal, we first pick the total number of events $N$ from a Poisson distribution 
with an expected total number of events $\langle N \rangle = 142.13$
(see the corresponding entry for $d=10$~kpc in Table~\ref{tab:predicted_event_numbers}),
and then draw $N$ events from the distribution $p(E,t|\text{z9.6-LS220}{\rm (NO)})$
[see Eq.~(\ref{eq:etdistr})] to form a set $\{E_i,t_i|i=1,2,\cdots,N\}$, where 
$E_i$ and $t_i$ are the energy and emission time of the $i$th event. Finally, 
for practical purposes, we define the detection time for the $i$th event as 
$t_{{\rm det},i}=t_i-t_1$ so that the first detected event corresponds to 
$t_{\rm det}=0$. We denote each simulated signal by the data set 
$D=\{E_i,t_{{\rm det},i}|i=1,2,\cdots,N\}$. 

Clearly, to compare a specific model $M_\alpha$ with the data, we need to introduce
a time offset $t_{\rm off}$ between $t=0$ for the start of neutrino emission in the model and 
$t_{\rm det}=0$ for detection of the first event so that $t=t_{\rm det}+t_{\rm off}$
(e.g., \cite{Olsen2021, Loredo:2001rx}; the time of travel from the SN to the detector 
is the same for all the events, and therefore, can be ignored).
The Bayesian approach to model comparison dictates that for $M_\alpha$ and the data $D$, 
the relevant quantity is the Bayesian evidence 
\begin{equation}
    P(D|M_\alpha) = \int dt_{\rm off} P(D|t_{\rm off}, M_\alpha) P(t_{\rm off}),
\end{equation}
where $P(D|t_{\rm off}, M_\alpha)$ is the likelihood function 
assuming $M_{\alpha}$, and $P(t_{\rm off})$ is the prior probability for $t_{\rm off}$. 
We take the likelihood of a simulated signal to be
\begin{equation}
    P(D|t_{\rm off}, M_\alpha) = \frac{e^{-\langle N \rangle} \langle N \rangle^N}{N!} 
    \prod_{i=1}^{N} p(E_i, t_i|t_{\rm off}, M_\alpha),
\end{equation}
which follows from the extended maximum likelihood function of \cite{BARLOW1990496}. We take 
the prior $P(t_{\rm off})$ to be uniform over the range $(0,0.1~\rm s)$ and $0$ otherwise. 

The Bayes factor
\begin{equation}
    B_{\alpha\beta} = \frac{P(D|M_\alpha)}{P(D|M_\beta)}
\end{equation}
can be used to determine whether $M_\alpha$ is favored over $M_\beta$ given the data $D$.
For convenience, we use the natural logarithm of the Bayes factor, $\ln B_{\alpha\beta}$, 
and refer to it simply as the Bayes factor. The criteria for interpreting 
$\ln B_{\alpha\beta}$ are shown in Table \ref{tab:ll_bayes_factor_interp} 
(e.g., \cite{Loredo:2001rx}). The larger $\ln B_{\alpha\beta}$ is, the more strongly 
$M_\alpha$ is favored over $M_\beta$.

\begin{table}
\caption{Interpretation of the Bayes factor $\ln B_{\alpha\beta}$
(partially reproduced from \cite{Loredo:2001rx}).}
\label{tab:ll_bayes_factor_interp}
\begin{tabular}{lc}
\hline 
\hline
$\ln B_{\alpha\beta}$ & Strength of Evidence \\
\hline
0--1 & Not worth more than a bare mention \\
1--3 & Positive\\
3--5 & Strong\\
$>5$ & Very strong\\
\hline 
\hline
\end{tabular}
\end{table}

We calculate the Bayesian evidence for each of our models, and compute the Bayes factors 
with $M_\alpha$ and $M_\beta$ being z9.6-LS220(NO) and each of the corresponding alternative 
models, respectively. Performing this procedure for the $10^4$ simulated
signals in our Monte Carlo sample allows us to calculate the mean 
$\langle \ln B_{\alpha\beta} \rangle$ and standard deviation $\sigma[\ln B_{\alpha\beta}]$ 
for the six model pairs. The results are displayed in the second column of Table~\ref{tab:unknown_offset}.
We see that for each alternative model, $\langle \ln B_{\alpha\beta} \rangle > 5$,
which indicates that at $d=50$~kpc with an ideal detector, a neutrino signal following z9.6-LS220(NO) 
would on average provide very strong evidence in favor of the true model over the corresponding alternatives.

\begin{table}
\caption{The mean Bayes factors along with standard deviations 
$\langle \ln B_{\alpha\beta} \rangle\pm\sigma[\ln B_{\alpha\beta}]$
calculated with $M_\alpha$ being z9.6-LS220(NO) using $10^4$ simulated signals in the assumed ideal 
detector from an SN at $d=50$~kpc. The true model $M_\alpha$ can be distinguished from the alternative 
$M_\beta$ at the $>95\%$ CL for $\langle \ln B_{\alpha\beta} \rangle - 1.645\sigma[\ln B_{\alpha\beta}]>5$,
which is satisfied by all the entries except for those in bold.}
\label{tab:unknown_offset}
\begin{tabular}{lcc}
    \hline\hline
    $M_\beta$ & Unknown $t_{\rm off}$ & $t_{\rm off}=t_1$\\
    \hline 
    z9.6-SFHo(NO) & $\mathbf{9.52 \pm 3.81}$ & $\mathbf{9.55 \pm 3.81}$ \\
    s18.6-LS220(NO) & $\mathbf{10.61 \pm 4.17}$ & $\mathbf{10.80 \pm 4.21}$ \\
    s18.6-SFHo(NO) & $21.81 \pm 5.32$ &  $21.97 \pm 5.35$\\
    s20-SFHo(NO) & $132.98 \pm 11.54$ &  $133.42 \pm 11.56 $\\
    s27-LS220(NO) & $70.14 \pm 9.12$ & $70.68 \pm 9.15$ \\
    s27-SFHo(NO) & $74.50 \pm 9.18$ & $74.82 \pm 9.20$ \\
    \hline\hline 
\end{tabular}
\end{table}

\begin{figure}
\includegraphics[width=\columnwidth]{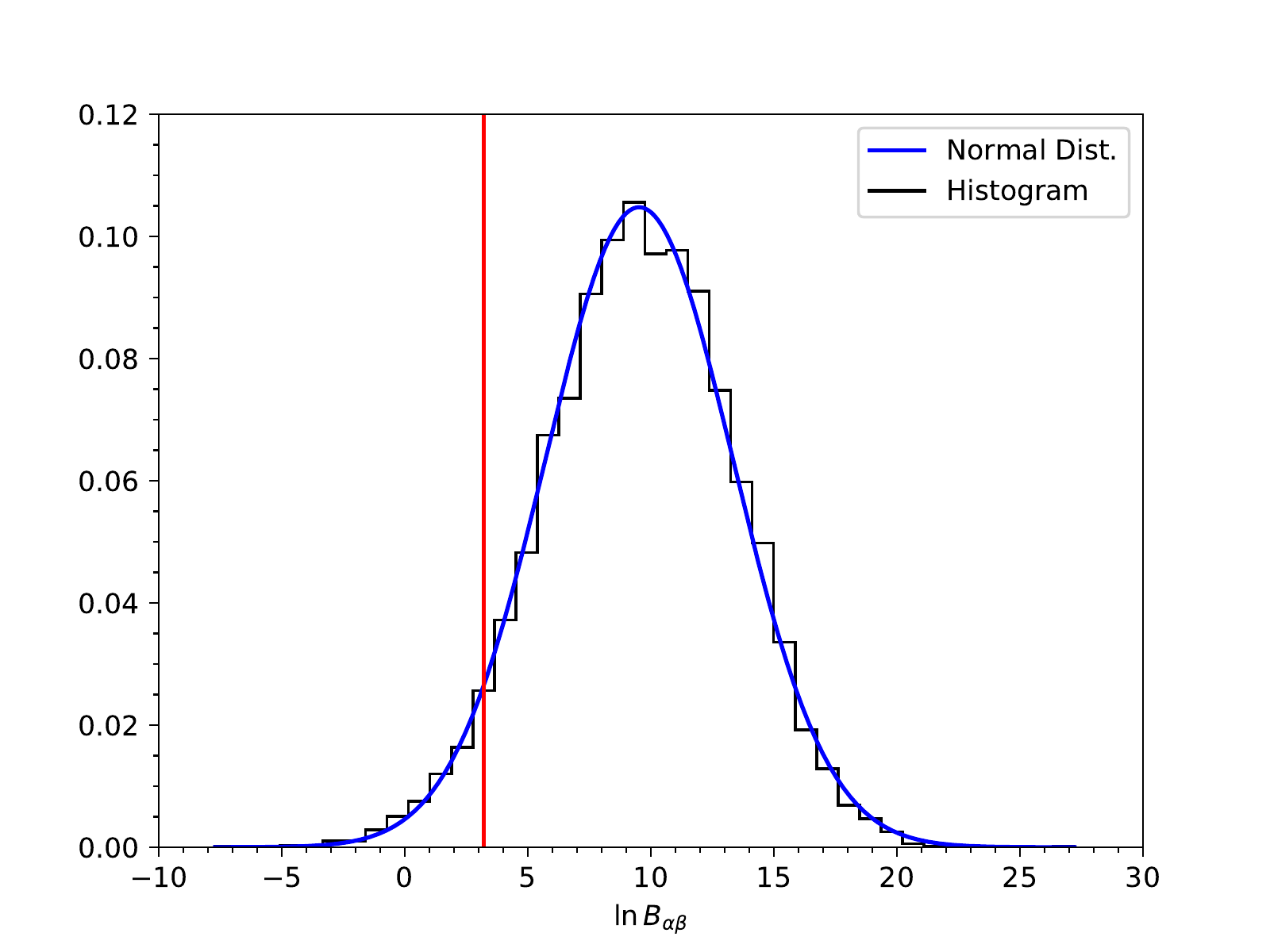}
\caption{The histogram of $\ln B_{\alpha\beta}$ is compared to the curve for the corresponding 
normal distribution with $M_\alpha$ and $M_\beta$ being z9.6-LS220(NO) and z9.6-SFHo(NO), 
respectively. The vertical line corresponds to 
$\ln B_{\alpha\beta}=\langle \ln B_{\alpha\beta} \rangle-1.645\sigma[\ln B_{\alpha\beta}]$,
and 95\% of the $\ln B_{\alpha\beta}$ values lie to the right of this line.
These results are based on $10^4$ simulated signals in the assumed ideal detector
from an SN at $d=50$~kpc. An unknown time offset $t_{\rm off}$ between the start of neutrino
emission and detection of the first event is taken into account.}
\label{fig:hist}
\end{figure}

We can set an even more stringent criterion for model distinguishability by noting 
that the distribution of $\ln B_{\alpha\beta}$ is approximately normal as demonstrated
in Fig.~\ref{fig:hist} with $M_\alpha$ and $M_\beta$ being z9.6-LS220(NO) and 
z9.6-SFHo(NO), respectively. The histogram in Fig.~\ref{fig:hist} shows the binned
results for the Monte Carlo sample, 
which are very well described by the curve for the normal distribution with 
the corresponding $\langle \ln B_{\alpha\beta} \rangle$ and $\sigma[\ln B_{\alpha\beta}]$.
We consider that the true model $M_\alpha$ can be distinguished from the alternative 
$M_\beta$ for $\ln B_{\alpha\beta}>5$.
We say that $M_\alpha$ and $M_\beta$ are distinguishable at the $>95\%$ confidence level (CL) if $\langle \ln B_{\alpha\beta} \rangle - 1.645\sigma[\ln B_{\alpha\beta}]>5$.
From the $\langle \ln B_{\alpha\beta} \rangle$ and $\sigma[\ln B_{\alpha\beta}]$ values
in the second column of Table~\ref{tab:unknown_offset}, we see that at $d=50$~kpc with an 
ideal detector, a neutrino signal following z9.6-LS220(NO) can be distinguished at the 
$>95\%$ CL from all the other adopted models except for z9.6-SFHo(NO) (see also 
Fig.~\ref{fig:hist}) and s18.6-LS220(NO). 

\subsection{\label{distances}General Results}
While the procedure in Sec.~\ref{montecarlo} for comparing $M_\alpha$ and $M_\beta$ 
is straightforward, it can be simplified by setting $t_{\rm off}=t_1$, where $t_1$ 
is the emission time of the first event for each simulated signal of $M_\alpha$. 
With this simplification, we are directly comparing the distributions of neutrino energy and 
emission time for $M_\alpha$ and $M_\beta$ by ignoring the effects of the time offset $t_{\rm off}$
between the start of emission and detection of the first event, and the Bayesian evidence for 
$M_\beta$ is simply
\begin{equation}
    P(D|M_\beta) = \frac{e^{-\langle N \rangle} \langle N \rangle^N}{N!} \prod_{i=1}^{N} p(E_i, t_i|M_\beta),
\end{equation}
where $\{E_i,t_i|i=1,2,\cdots,N\}$ effectively denotes a simulated signal of $M_\alpha$.
Using the $10^4$ simulated signals of z9.6-LS220(NO) ($M_\alpha$) in Sec.~\ref{montecarlo} and setting
$t_{\rm off}=t_1$, we calculate the $\langle \ln B_{\alpha\beta} \rangle$ and $\sigma[\ln B_{\alpha\beta}]$ 
values for comparing this model with all the corresponding alternatives. As can be seen from
Table~\ref{tab:unknown_offset}, these results are approximately the same as those that have taken the
effects of $t_{\rm off}$ into account. Therefore, we can ignore the very small effects of $t_{\rm off}$ 
in calculating $\langle \ln B_{\alpha\beta} \rangle$ and $\sigma[\ln B_{\alpha\beta}]$ to determine the
distinguishability of $M_\alpha$ and $M_\beta$.

Below we compare all pairs of our adopted models using the simplified procedure that does not include 
$t_{\rm off}$ as a parameter. The corresponding Bayes factor is
\begin{equation}
    \ln B_{\alpha\beta} = N\ln\frac{\langle N \rangle_\alpha}{\langle N \rangle_\beta} - \Delta_{\alpha\beta} 
    + \sum_{i=1}^{N} \ln \frac{p(E_i, t_i|M_\alpha)}{p(E_i, t_i|M_\beta)},
\end{equation}
where $\langle N \rangle_\alpha$ and $\langle N \rangle_\beta$ are the expected total numbers of events 
predicted by $M_\alpha$ and $M_\beta$, respectively, for an SN at a known distance $d$, and 
$\Delta_{\alpha\beta} \equiv \langle N \rangle_\alpha - \langle N \rangle_\beta$.
With simulated neutrino signals from $M_\alpha$, $N$ is sampled from the Poisson distribution with the mean 
$\langle N \rangle_\alpha$, and $\{E_i,t_i|i=1,2,\cdots,N\}$ is sampled from the energy-time distribution 
$\prod_{j=1}^{N} p(E_j, t_j|M_\alpha)$. Instead of Monte Carlo simulations, we can use the above 
distributions directly to obtain
\begin{equation}
\label{eq:bf_mean_eml}
    \frac{\langle \ln B_{\alpha\beta} \rangle}{\langle N \rangle_\alpha}= 
    \ln\frac{\langle N \rangle_\alpha}{\langle N \rangle_\beta} - 
    \frac{\Delta_{\alpha\beta}}{\langle N \rangle_\alpha} + 
    \left\langle \ln \frac{p(E,t|M_\alpha)}{p(E,t|M_\beta)} \right\rangle_\alpha
\end{equation}
and
\begin{equation}
\label{eq:bf_std_eml}
\begin{split}
    \frac{\sigma[\ln B_{\alpha\beta}]}{\sqrt{\langle N \rangle_\alpha}}& = \Bigg[
    \left(\ln\frac{\langle N \rangle_\alpha}{\langle N \rangle_\beta}\right)^2\\ 
    &+ 2 \ln\frac{\langle N \rangle_\alpha}{\langle N \rangle_\beta} 
    \left\langle \ln\frac{p(E, t|M_\alpha)}{p(E, t|M_\beta)} \right\rangle_\alpha\\ 
    &+ \left\langle \left(\ln\frac{p(E, t|M_\alpha)}{p(E, t|M_\beta)}\right)^2 \right\rangle_\alpha
    \Bigg]^{1/2},
\end{split}
\end{equation}
where, for example,
\begin{equation}
\begin{split}
    \left\langle \ln \frac{p(E,t|M_\alpha)}{p(E,t|M_\beta)} \right\rangle_\alpha&=
    \int_0^{9\ {\rm s}} dt\int_{E_{\rm min}}^\infty dE\ p(E,t|M_\alpha)\\
    &\times\ln \frac{p(E,t|M_\alpha)}{p(E,t|M_\beta)}.
\end{split}
\end{equation}
Both $\langle \ln B_{\alpha\beta} \rangle$ and $\sigma[\ln B_{\alpha\beta}]$ are 
functions of the SN distance $d$ and depend on the assumed detector.

The $\langle \ln B_{\alpha\beta} \rangle$ and $\sigma[\ln B_{\alpha\beta}]$ values for
all pairs of the adopted SN neutrino emission models are given for an SN at $d=50$ (25)~kpc 
in Tables~\ref{tab:dist50_ideal} and \ref{tab:dist50_sk} (\ref{tab:dist25_ideal} and 
\ref{tab:dist25_sk}) for the assumed ideal and SK-like detectors, respectively.
We see that, for either detector, all of the models can be distinguished from 
one another at the $>95\%$ CL 
(with $\langle \ln B_{\alpha\beta} \rangle - 1.645\sigma[\ln B_{\alpha\beta}]>5$) 
for $d=25$~kpc, and some of the models can still be distinguished at the same CL 
for $d=50$~kpc. Note that the results for comparing z9.6-LS220(NO) ($M_\alpha$) with
the corresponding alternatives in Table~\ref{tab:dist50_ideal} are in excellent
agreement with those in Table~\ref{tab:unknown_offset} calculated with Monte Carlo simulations.

\begin{table*}
\caption{The mean Bayes factors along with standard deviations 
$\langle \ln B_{\alpha\beta} \rangle\pm\sigma[\ln B_{\alpha\beta}]$
calculated for the assumed ideal detector and an SN at $d=50$~kpc.
The true model $M_\alpha$ can be distinguished from the alternative $M_\beta$ at
the $>95\%$ CL for $\langle \ln B_{\alpha\beta} \rangle - 1.645\sigma[\ln B_{\alpha\beta}]>5$,
which is satisfied by all the entries except for those in bold.}
\label{tab:dist50_ideal}
\begin{tabular}{cccccccc}
    \hline\hline
    $M_\alpha$\textbackslash$M_\beta$  & z9.6-LS220 & z9.6-SFHo & s18.6-LS220 & s18.6-SFHo & s20-SFHo & s27-LS220 & s27-SFHo\\
    \hline 
z9.6-LS220&&&&&&&\\
\makecell{(NO)\\(NH)\\(IH)}& \makecell{} & \makecell{$\mathbf{9.53\pm3.86}$\\ $\mathbf{10.75\pm4.05}$\\ $13.70\pm4.52$} & \makecell{$\mathbf{10.78\pm4.23}$\\ $\mathbf{10.78\pm4.28}$\\ $\mathbf{11.41\pm4.43}$} & \makecell{$21.93\pm5.37$\\ $22.01\pm5.32$\\ $22.97\pm5.29$} & \makecell{$133.28\pm11.66$\\ $122.07\pm11.34$\\ $104.54\pm10.45$} & \makecell{$70.59\pm9.24$\\ $60.77\pm8.89$\\ $46.21\pm8.01$} & \makecell{$74.73\pm9.28$\\ $69.16\pm8.97$\\ $61.74\pm8.29$}\\ 
z9.6-SFHo&&&&&&&\\
\makecell{(NO)\\(NH)\\(IH)}& \makecell{$\mathbf{12.98\pm6.07}$\\ $\mathbf{15.00\pm6.60}$\\ $19.65\pm7.66$} & \makecell{} & \makecell{$25.58\pm7.22$\\ $29.13\pm7.85$\\ $37.53\pm9.12$} & \makecell{$15.28\pm4.97$\\ $14.86\pm4.95$\\ $14.63\pm4.96$} & \makecell{$125.61\pm11.78$\\ $113.4\pm11.61$\\ $92.84\pm11.00$} & \makecell{$75.15\pm9.18$\\ $65.89\pm8.99$\\ $51.74\pm8.41$} & \makecell{$67.69\pm9.26$\\ $61.10\pm9.07$\\ $50.82\pm8.57$}\\ 
s18.6-LS220&&&&&&&\\
\makecell{(NO)\\(NH)\\(IH)}& \makecell{$\mathbf{13.01\pm5.60}$\\ $\mathbf{12.68\pm5.46}$\\ $\mathbf{13.23\pm5.53}$} & \makecell{$27.86\pm8.08$\\ $30.75\pm8.38$\\ $38.34\pm9.27$} & \makecell{} & \makecell{$11.93\pm3.87$\\ $13.95\pm4.20$\\ $18.62\pm4.89$} & \makecell{$91.85\pm9.75$\\ $82.57\pm9.25$\\ $70.26\pm8.12$} & \makecell{$38.85\pm6.98$\\ $30.28\pm6.40$\\ $19.57\pm5.22$} & \makecell{$45.73\pm7.14$\\ $42.12\pm6.76$\\ $39.55\pm6.12$}\\ 
s18.6-SFHo&&&&&&&\\
\makecell{(NO)\\(NH)\\(IH)}& \makecell{$33.59\pm10.17$\\ $34.43\pm10.41$\\ $37.58\pm11.09$} & \makecell{$18.83\pm6.81$\\ $17.88\pm6.56$\\ $17.19\pm6.37$} & \makecell{$19.23\pm7.89$\\ $22.51\pm8.56$\\ $29.87\pm9.89$} & \makecell{} & \makecell{$71.02\pm9.32$\\ $60.49\pm8.93$\\ $44.02\pm7.95$} & \makecell{$37.30\pm6.77$\\ $29.20\pm6.37$\\ $18.30\pm5.50$} & \makecell{$28.55\pm6.24$\\ $23.16\pm5.84$\\ $15.71\pm5.01$}\\ 
s20-SFHo&&&&&&&\\
\makecell{(NO)\\(NH)\\(IH)}& \makecell{$258.83\pm31.60$\\ $232.91\pm29.89$\\ $206.21\pm28.78$} & \makecell{$225.48\pm28.40$\\ $187.49\pm24.74$\\ $140.18\pm20.47$} & \makecell{$180.31\pm26.75$\\ $164.28\pm25.83$\\ $153.87\pm26.04$} & \makecell{$117.55\pm19.80$\\ $92.27\pm16.80$\\ $62.25\pm13.34$} & \makecell{} & \makecell{$34.63\pm10.12$\\ $36.36\pm10.47$\\ $40.68\pm11.24$} & \makecell{$15.68\pm6.08$\\ $\mathbf{13.45\pm5.58}$\\ $\mathbf{9.85\pm4.72}$}\\ 
s27-LS220&&&&&&&\\
\makecell{(NO)\\(NH)\\(IH)}& \makecell{$115.89\pm19.44$\\ $92.48\pm16.70$\\ $66.25\pm13.76$} & \makecell{$134.20\pm21.90$\\ $106.55\pm18.50$\\ $75.20\pm14.75$} & \makecell{$61.80\pm14.00$\\ $44.63\pm11.45$\\ $28.29\pm9.06$} & \makecell{$61.91\pm14.40$\\ $42.89\pm11.28$\\ $22.89\pm7.63$} & \makecell{$23.44\pm5.65$\\ $24.10\pm5.66$\\ $26.01\pm5.76$} & \makecell{} & \makecell{$\mathbf{11.08\pm4.30}$\\ $\mathbf{11.26\pm4.22}$\\ $12.64\pm4.32$}\\ 
s27-SFHo&&&&&&&\\
\makecell{(NO)\\(NH)\\(IH)}& \makecell{$129.99\pm21.28$\\ $120.53\pm20.60$\\ $113.79\pm20.60$} & \makecell{$106.66\pm18.34$\\ $89.85\pm16.20$\\ $69.67\pm13.77$} & \makecell{$82.91\pm17.41$\\ $79.10\pm17.32$\\ $82.27\pm18.34$} & \makecell{$42.30\pm11.23$\\ $31.64\pm9.30$\\ $20.01\pm7.13$} & \makecell{$13.35\pm4.77$\\ $\mathbf{11.64\pm4.49}$\\ $\mathbf{8.73\pm3.94}$} & \makecell{$\mathbf{13.78\pm5.94}$\\ $\mathbf{14.66\pm6.24}$\\ $17.39\pm6.95$} & \makecell{}\\ 
\hline\hline 
\end{tabular}
\end{table*}

\begin{table*}
\caption{The mean Bayes factors along with standard deviations 
$\langle \ln B_{\alpha\beta} \rangle\pm\sigma[\ln B_{\alpha\beta}]$
calculated for the assumed SK-like detector and an SN at $d=50$~kpc.
The true model $M_\alpha$ can be distinguished from the alternative $M_\beta$ at
the $>95\%$ CL for $\langle \ln B_{\alpha\beta} \rangle - 1.645\sigma[\ln B_{\alpha\beta}]>5$,
which is satisfied by all the entries except for those in bold.}
\label{tab:dist50_sk}
\begin{tabular}{cccccccc}
\hline\hline
$M_\alpha$\textbackslash$M_\beta$ & z9.6-LS220 & z9.6-SFHo & s18.6-LS220 & s18.6-SFHo & s20-SFHo & s27-LS220 & s27-SFHo\\
\hline
z9.6-LS220&&&&&&&\\ 
\makecell{(NO)\\(NH)\\(IH)}& \makecell{} & \makecell{$\mathbf{6.88\pm3.27}$\\ $\mathbf{7.80\pm3.45}$\\ $\mathbf{9.98\pm3.86}$} & \makecell{$\mathbf{7.88\pm3.62}$\\ $\mathbf{7.91\pm3.66}$\\ $\mathbf{8.42\pm3.81}$} & \makecell{$15.96\pm4.58$\\ $16.04\pm4.54$\\ $16.75\pm4.52$} & \makecell{$97.83\pm10.00$\\ $89.56\pm9.72$\\ $76.54\pm8.95$} & \makecell{$51.96\pm7.94$\\ $44.71\pm7.64$\\ $33.91\pm6.88$} & \makecell{$54.73\pm7.95$\\ $50.65\pm7.68$\\ $45.16\pm7.09$}\\ 
z9.6-SFHo&&&&&&&\\
\makecell{(NO)\\(NH)\\(IH)}& \makecell{$\mathbf{9.36\pm5.14}$\\ $\mathbf{10.85\pm5.59}$\\ $\mathbf{14.24\pm6.49}$} & \makecell{} & \makecell{$18.71\pm6.15$\\ $21.34\pm6.69$\\ $27.49\pm7.77$} & \makecell{$\mathbf{11.22\pm4.26}$\\ $\mathbf{10.94\pm4.25}$\\ $\mathbf{10.78\pm4.26}$} & \makecell{$92.54\pm10.14$\\ $83.55\pm9.98$\\ $68.33\pm9.45$} & \makecell{$55.49\pm7.93$\\ $48.69\pm7.76$\\ $38.23\pm7.25$} & \makecell{$49.82\pm7.97$\\ $45.00\pm7.80$\\ $37.41\pm7.36$}\\ 
s18.6-LS220&&&&&&&\\
\makecell{(NO)\\(NH)\\(IH)}& \makecell{$\mathbf{9.50\pm4.79}$\\ $\mathbf{9.30\pm4.67}$\\ $\mathbf{9.75\pm4.75}$} & \makecell{$20.37\pm6.89$\\ $22.56\pm7.17$\\ $28.18\pm7.94$} & \makecell{} & \makecell{$\mathbf{8.72\pm3.31}$\\ $\mathbf{10.20\pm3.59}$\\ $13.61\pm4.19$} & \makecell{$67.47\pm8.37$\\ $60.55\pm7.93$\\ $51.28\pm6.94$} & \makecell{$28.60\pm6.02$\\ $22.24\pm5.50$\\ $14.23\pm4.46$} & \makecell{$33.52\pm6.13$\\ $30.83\pm5.79$\\ $28.83\pm5.23$}\\ 
s18.6-SFHo&&&&&&&\\
\makecell{(NO)\\(NH)\\(IH)}& \makecell{$24.37\pm8.64$\\ $24.99\pm8.84$\\ $27.27\pm9.41$} & \makecell{$\mathbf{13.79\pm5.82}$\\ $\mathbf{13.13\pm5.61}$\\ $\mathbf{12.64\pm5.46}$} & \makecell{$\mathbf{13.95\pm6.68}$\\ $\mathbf{16.33\pm7.25}$\\ $21.66\pm8.38$} & \makecell{} & \makecell{$52.31\pm8.02$\\ $44.53\pm7.68$\\ $32.29\pm6.82$} & \makecell{$27.51\pm5.85$\\ $21.54\pm5.49$\\ $13.44\pm4.72$} & \makecell{$21.01\pm5.37$\\ $17.03\pm5.02$\\ $\mathbf{11.50\pm4.29}$}\\ 
s20-SFHo&&&&&&&\\
\makecell{(NO)\\(NH)\\(IH)}& \makecell{$188.41\pm26.79$\\ $169.63\pm25.35$\\ $149.66\pm24.33$} & \makecell{$164.42\pm24.08$\\ $137.31\pm21.09$\\ $102.66\pm17.46$} & \makecell{$130.95\pm22.61$\\ $119.22\pm21.82$\\ $111.11\pm21.92$} & \makecell{$85.72\pm16.79$\\ $67.5\pm14.31$\\ $45.40\pm11.35$} & \makecell{} & \makecell{$25.25\pm8.62$\\ $26.50\pm8.92$\\ $29.63\pm9.57$} & \makecell{$\mathbf{11.51\pm5.20}$\\ $\mathbf{9.88\pm4.77}$\\ $\mathbf{7.22\pm4.04}$}\\ 
s27-LS220&&&&&&&\\
\makecell{(NO)\\(NH)\\(IH)}& \makecell{$84.57\pm16.52$\\ $67.67\pm14.24$\\ $48.35\pm11.71$} & \makecell{$97.67\pm18.52$\\ $78.02\pm15.75$\\ $55.14\pm12.58$} & \makecell{$44.99\pm11.87$\\ $32.54\pm9.73$\\ $20.43\pm7.67$} & \makecell{$45.01\pm12.18$\\ $31.35\pm9.59$\\ $16.68\pm6.48$} & \makecell{$17.13\pm4.83$\\ $17.61\pm4.84$\\ $19.01\pm4.92$} & \makecell{} & \makecell{$\mathbf{8.05\pm3.67}$\\ $\mathbf{8.21\pm3.60}$\\ $\mathbf{9.24\pm3.70}$}\\ 
s27-SFHo&&&&&&&\\
\makecell{(NO)\\(NH)\\(IH)}& \makecell{$94.60\pm18.06$\\ $87.72\pm17.48$\\ $82.57\pm17.44$} & \makecell{$77.85\pm15.58$\\ $65.85\pm13.83$\\ $51.07\pm11.76$} & \makecell{$60.17\pm14.72$\\ $57.33\pm14.63$\\ $59.38\pm15.46$} & \makecell{$30.85\pm9.54$\\ $23.15\pm7.93$\\ $\mathbf{14.58\pm6.07}$} & \makecell{$\mathbf{9.82\pm4.10}$\\ $\mathbf{8.56\pm3.86}$\\ $\mathbf{6.41\pm3.37}$} & \makecell{$\mathbf{10.02\pm5.06}$\\ $\mathbf{10.67\pm5.31}$\\ $\mathbf{12.67\pm5.92}$} & \makecell{}\\ 
\hline\hline 
\end{tabular}
\end{table*} 

\begin{table*}
\caption{The mean Bayes factors along with standard deviations 
$\langle \ln B_{\alpha\beta} \rangle\pm\sigma[\ln B_{\alpha\beta}]$
calculated for the assumed ideal detector and an SN at $d=25$~kpc.
The true model $M_\alpha$ can be distinguished from the alternative $M_\beta$ at
the $>95\%$ CL for $\langle \ln B_{\alpha\beta} \rangle - 1.645\sigma[\ln B_{\alpha\beta}]>5$,
which is satisfied by all the entries.}
\label{tab:dist25_ideal}
\begin{tabular}{cccccccc}
    \hline\hline
$M_\alpha$\textbackslash$M_\beta$ & z9.6-LS220 & z9.6-SFHo & s18.6-LS220 & s18.6-SFHo & s20-SFHo & s27-LS220 & s27-SFHo\\
    \hline 
z9.6-LS220&&&&&&&\\
\makecell{(NO)\\(NH)\\(IH)}& \makecell{} & \makecell{$38.13\pm7.72$\\ $42.99\pm8.10$\\ $54.82\pm9.03$} & \makecell{$43.14\pm8.46$\\ $43.12\pm8.56$\\ $45.64\pm8.86$} & \makecell{$87.74\pm10.75$\\ $88.05\pm10.64$\\ $91.88\pm10.57$} & \makecell{$533.14\pm23.32$\\ $488.30\pm22.68$\\ $418.17\pm20.91$} & \makecell{$282.38\pm18.47$\\ $243.07\pm17.77$\\ $184.85\pm16.02$} & \makecell{$298.91\pm18.55$\\ $276.64\pm17.94$\\ $246.98\pm16.58$}\\ 
z9.6-SFHo&&&&&&&\\
\makecell{(NO)\\(NH)\\(IH)}& \makecell{$51.94\pm12.15$\\ $60.00\pm13.20$\\ $78.61\pm15.32$} & \makecell{} & \makecell{$102.30\pm14.44$\\ $116.54\pm15.70$\\ $150.13\pm18.25$} & \makecell{$61.11\pm9.94$\\ $59.46\pm9.90$\\ $58.52\pm9.91$} & \makecell{$502.43\pm23.56$\\ $453.60\pm23.22$\\ $371.34\pm22.01$} & \makecell{$300.59\pm18.36$\\ $263.55\pm17.98$\\ $206.95\pm16.82$} & \makecell{$270.75\pm18.53$\\ $244.40\pm18.15$\\ $203.28\pm17.14$}\\ 
s18.6-LS220&&&&&&&\\
\makecell{(NO)\\(NH)\\(IH)}& \makecell{$52.03\pm11.21$\\ $50.7\pm10.91$\\ $52.92\pm11.06$} & \makecell{$111.46\pm16.15$\\ $122.98\pm16.75$\\ $153.37\pm18.54$} & \makecell{} & \makecell{$47.7\pm7.74$\\ $55.82\pm8.39$\\ $74.47\pm9.79$} & \makecell{$367.40\pm19.50$\\ $330.26\pm18.50$\\ $281.05\pm16.24$} & \makecell{$155.42\pm13.96$\\ $121.12\pm12.80$\\ $78.28\pm10.43$} & \makecell{$182.91\pm14.29$\\ $168.47\pm13.51$\\ $158.20\pm12.24$}\\ 
s18.6-SFHo&&&&&&&\\
\makecell{(NO)\\(NH)\\(IH)}& \makecell{$134.35\pm20.34$\\ $137.71\pm20.81$\\ $150.33\pm22.19$} & \makecell{$75.32\pm13.62$\\ $71.52\pm13.12$\\ $68.76\pm12.75$} & \makecell{$76.94\pm15.77$\\ $90.04\pm17.12$\\ $119.47\pm19.78$} & \makecell{} & \makecell{$284.09\pm18.64$\\ $241.96\pm17.87$\\ $176.10\pm15.91$} & \makecell{$149.18\pm13.55$\\ $116.79\pm12.74$\\ $73.20\pm10.99$} & \makecell{$114.19\pm12.48$\\ $92.65\pm11.69$\\ $62.85\pm10.01$}\\ 
s20-SFHo&&&&&&&\\
\makecell{(NO)\\(NH)\\(IH)}& \makecell{$1035.31\pm63.20$\\ $931.65\pm59.78$\\ $824.83\pm57.56$} & \makecell{$901.91\pm56.79$\\ $749.94\pm49.49$\\ $560.73\pm40.94$} & \makecell{$721.24\pm53.50$\\ $657.10\pm51.66$\\ $615.50\pm52.07$} & \makecell{$470.21\pm39.59$\\ $369.08\pm33.59$\\ $249.00\pm26.67$} & \makecell{} & \makecell{$138.53\pm20.25$\\ $145.45\pm20.94$\\ $162.73\pm22.48$} & \makecell{$62.73\pm12.15$\\ $53.79\pm11.15$\\ $39.41\pm9.44$}\\ 
s27-LS220&&&&&&&\\
\makecell{(NO)\\(NH)\\(IH)}& \makecell{$463.55\pm38.88$\\ $369.91\pm33.40$\\ $264.98\pm27.51$} & \makecell{$536.80\pm43.79$\\ $426.20\pm37.00$\\ $300.79\pm29.51$} & \makecell{$247.19\pm27.99$\\ $178.54\pm22.90$\\ $113.16\pm18.12$} & \makecell{$247.63\pm28.81$\\ $171.55\pm22.55$\\ $91.56\pm15.25$} & \makecell{$93.76\pm11.30$\\ $96.42\pm11.32$\\ $104.05\pm11.51$} & \makecell{} & \makecell{$44.31\pm8.61$\\ $45.05\pm8.45$\\ $50.57\pm8.65$}\\ 
s27-SFHo&&&&&&&\\
\makecell{(NO)\\(NH)\\(IH)}& \makecell{$519.96\pm42.57$\\ $482.12\pm41.20$\\ $455.17\pm41.21$} & \makecell{$426.63\pm36.68$\\ $359.41\pm32.40$\\ $278.68\pm27.54$} & \makecell{$331.64\pm34.82$\\ $316.41\pm34.63$\\ $329.06\pm36.68$} & \makecell{$169.21\pm22.46$\\ $126.58\pm18.60$\\ $80.04\pm14.26$} & \makecell{$53.38\pm9.54$\\ $46.57\pm8.98$\\ $34.92\pm7.87$} & \makecell{$55.13\pm11.89$\\ $58.63\pm12.48$\\ $69.56\pm13.89$} & \makecell{}\\ 
\hline\hline
\end{tabular}
\end{table*} 

\begin{table*}
\caption{The mean Bayes factors along with standard deviations 
$\langle \ln B_{\alpha\beta} \rangle\pm\sigma[\ln B_{\alpha\beta}]$
calculated for the assumed SK-like detector and an SN at $d=25$~kpc.
The true model $M_\alpha$ can be distinguished from the alternative $M_\beta$ at
the $>95\%$ CL for $\langle \ln B_{\alpha\beta} \rangle - 1.645\sigma[\ln B_{\alpha\beta}]>5$,
which is satisfied by all the entries.}
\label{tab:dist25_sk}
\begin{tabular}{cccccccc}
\hline\hline
$M_\alpha$\textbackslash$M_\beta$  & z9.6-LS220 & z9.6-SFHo & s18.6-LS220 & s18.6-SFHo & s20-SFHo & s27-LS220 & s27-SFHo\\
\hline 
z9.6-LS220&&&&&&&\\
\makecell{(NO)\\(NH)\\(IH)}& \makecell{} & \makecell{$27.53\pm6.55$\\ $31.21\pm6.89$\\ $39.94\pm7.72$} & \makecell{$31.53\pm7.23$\\ $31.64\pm7.33$\\ $33.68\pm7.62$} & \makecell{$63.84\pm9.16$\\ $64.14\pm9.08$\\ $67.01\pm9.03$} & \makecell{$391.32\pm20.00$\\ $358.23\pm19.43$\\ $306.16\pm17.90$} & \makecell{$207.82\pm15.89$\\ $178.84\pm15.27$\\ $135.65\pm13.75$} & \makecell{$218.92\pm15.89$\\ $202.58\pm15.36$\\ $180.62\pm14.19$}\\ 
z9.6-SFHo&&&&&&&\\
\makecell{(NO)\\(NH)\\(IH)}& \makecell{$37.46\pm10.29$\\ $43.40\pm11.19$\\ $56.94\pm12.98$} & \makecell{} & \makecell{$74.85\pm12.31$\\ $85.37\pm13.38$\\ $109.97\pm15.54$} & \makecell{$44.87\pm8.53$\\ $43.74\pm8.50$\\ $43.11\pm8.51$} & \makecell{$370.16\pm20.28$\\ $334.21\pm19.96$\\ $273.31\pm18.91$} & \makecell{$221.95\pm15.86$\\ $194.77\pm15.52$\\ $152.91\pm14.50$} & \makecell{$199.28\pm15.93$\\ $179.99\pm15.60$\\ $149.65\pm14.73$}\\ 
s18.6-LS220&&&&&&&\\
\makecell{(NO)\\(NH)\\(IH)}& \makecell{$38.00\pm9.57$\\ $37.19\pm9.34$\\ $39.01\pm9.49$} & \makecell{$81.50\pm13.78$\\ $90.23\pm14.33$\\ $112.73\pm15.89$} & \makecell{} & \makecell{$34.87\pm6.63$\\ $40.81\pm7.19$\\ $54.43\pm8.38$} & \makecell{$269.86\pm16.75$\\ $242.20\pm15.86$\\ $205.12\pm13.88$} & \makecell{$114.4\pm12.03$\\ $88.97\pm11.00$\\ $56.90\pm8.91$} & \makecell{$134.10\pm12.26$\\ $123.32\pm11.58$\\ $115.33\pm10.46$}\\ 
s18.6-SFHo&&&&&&&\\
\makecell{(NO)\\(NH)\\(IH)}& \makecell{$97.48\pm17.27$\\ $99.96\pm17.67$\\ $109.06\pm18.82$} & \makecell{$55.14\pm11.63$\\ $52.52\pm11.23$\\ $50.56\pm10.92$} & \makecell{$55.80\pm13.36$\\ $65.31\pm14.50$\\ $86.65\pm16.76$} & \makecell{} & \makecell{$209.25\pm16.04$\\ $178.11\pm15.36$\\ $129.15\pm13.65$} & \makecell{$110.04\pm11.70$\\ $86.15\pm10.98$\\ $53.74\pm9.45$} & \makecell{$84.02\pm10.74$\\ $68.13\pm10.04$\\ $45.99\pm8.58$}\\ 
s20-SFHo&&&&&&&\\
\makecell{(NO)\\(NH)\\(IH)}& \makecell{$753.64\pm53.57$\\ $678.53\pm50.69$\\ $598.66\pm48.67$} & \makecell{$657.68\pm48.15$\\ $549.23\pm42.18$\\ $410.65\pm34.92$} & \makecell{$523.8\pm45.21$\\ $476.86\pm43.63$\\ $444.43\pm43.85$} & \makecell{$342.86\pm33.58$\\ $270.00\pm28.62$\\ $181.61\pm22.69$} & \makecell{} & \makecell{$100.98\pm17.25$\\ $106.00\pm17.83$\\ $118.51\pm19.13$} & \makecell{$46.04\pm10.40$\\ $39.51\pm9.55$\\ $28.89\pm8.07$}\\ 
s27-LS220&&&&&&&\\
\makecell{(NO)\\(NH)\\(IH)}& \makecell{$338.28\pm33.03$\\ $270.70\pm28.47$\\ $193.39\pm23.43$} & \makecell{$390.66\pm37.03$\\ $312.09\pm31.50$\\ $220.54\pm25.15$} & \makecell{$179.96\pm23.73$\\ $130.17\pm19.47$\\ $81.73\pm15.33$} & \makecell{$180.02\pm24.36$\\ $125.40\pm19.18$\\ $66.74\pm12.96$} & \makecell{$68.50\pm9.66$\\ $70.45\pm9.68$\\ $76.02\pm9.84$} & \makecell{} & \makecell{$32.21\pm7.33$\\ $32.84\pm7.21$\\ $36.96\pm7.39$}\\ 
s27-SFHo&&&&&&&\\
\makecell{(NO)\\(NH)\\(IH)}& \makecell{$378.39\pm36.12$\\ $350.88\pm34.96$\\ $330.26\pm34.88$} & \makecell{$311.41\pm31.16$\\ $263.39\pm27.65$\\ $204.29\pm23.52$} & \makecell{$240.68\pm29.44$\\ $229.34\pm29.26$\\ $237.52\pm30.92$} & \makecell{$123.41\pm19.07$\\ $92.61\pm15.86$\\ $58.31\pm12.14$} & \makecell{$39.28\pm8.19$\\ $34.26\pm7.71$\\ $25.63\pm6.75$} & \makecell{$40.08\pm10.12$\\ $42.68\pm10.63$\\ $50.69\pm11.83$} & \makecell{}\\ 
\hline\hline 
\end{tabular}
\end{table*} 

\subsection{Scenarios of Neutrino Oscillations}
So far we have assumed that the scenario of neutrino oscillations would be known
and whichever it is, it applies to all SN neutrino emission models in the same way.
It is interesting to note that the distinguishability of these models depends on the scenario
of neutrino oscillations. For example, for an SN at $d=50$~kpc with the assumed ideal detector,
we can distinguish z9.6-LS220(IH) and z9.6-SFHo(IH) at the $>95\%$ CL, but cannot do the same
for z9.6-LS220(NH) and z9.6-SFHo(NH) (see Table~\ref{tab:dist50_ideal}).

We now explore the feasibility of distinguishing between the scenarios of neutrino oscillations
for a specific underlying SN neutrino emission model. We again perform the comparisons using 
Eqs.~(\ref{eq:bf_mean_eml}) and (\ref{eq:bf_std_eml}).
With more significant differences between $\bar\nu_e$ and $\bar\nu_x$ emission during the
accretion phase, we expect that more pronounced accretion-induced emission allows for easier 
distinguishability of the oscillation scenarios. Because z9.6-LS220 and s20-SFHo represent 
the opposite extremes of accretion-induced neutrino emission (see Fig.~\ref{fig:models}),
we focus on these two models. The results for an SN at $d=50$, 25, and 10~kpc
with the assumed ideal detector are presented in Tables~\ref{tab:osc_z9.6} and \ref{tab:tab:osc_s20}.
We see that at $d=50$~kpc, only the (NO) and (IH) scenarios can be distinguished at the $>95\%$ CL
for s20-SFHo, but none of the scenarios can be distinguished for z9.6-LS220. 
At $d=25$~kpc, only the (NO) and (NH) scenarios cannot be distinguished for s20-SFHo, while only 
the (NO) and (IH) scenarios can be distinguished for z9.6-LS220.
Finally, at $d=10$~kpc, all oscillation scenarios can be distinguished for either model.

\begin{table}
\caption{The mean Bayes factors along with standard deviations 
$\langle \ln B_{\alpha\beta} \rangle\pm\sigma[\ln B_{\alpha\beta}]$
for comparing different scenarios of neutrino oscillations for
the same underlying SN neutrino emission model z9.6-LS220.
Three SN distances $d=50$, 25, and 10~kpc are used and the ideal detector is assumed.
The true model $M_\alpha$ can be distinguished from the alternative $M_\beta$ at
the $>95\%$ CL for $\langle \ln B_{\alpha\beta} \rangle - 1.645\sigma[\ln B_{\alpha\beta}]>5$,
which is satisfied by all the entries except for those in bold.}
\label{tab:osc_z9.6}
\begin{tabular}{cccc}
    \hline\hline
    $M_\alpha$\textbackslash$M_\beta$ & (NO) & (NH) & (IH)\\ 
    \hline
    $d=50$~kpc&&&\\
    (NO) &  & $\mathbf{1.10\pm 1.35}$ & $\mathbf{7.48\pm 3.44}$\\ 
    (NH) & $\mathbf{1.37\pm 1.89}$ &  & $\mathbf{3.15\pm 2.38}$\\ 
    (IH) & $\mathbf{9.98\pm 5.31}$ & $\mathbf{3.53\pm 2.82}$ & \\ 
    $d=25$~kpc&&&\\
    (NO) &  & $\mathbf{4.39\pm 2.71}$ & $29.92\pm 6.88$\\ 
    (NH) & $\mathbf{5.46\pm 3.77}$ & & $\mathbf{12.60\pm 4.76}$\\ 
    (IH) & $39.94\pm 10.61$ & $\mathbf{14.11\pm 5.64}$ & \\ 
    $d=10$~kpc&&&\\
    (NO) &  & $27.44\pm 6.77$ & $186.98\pm 17.21$\\ 
    (NH) & $34.15\pm 9.43$ &  & $78.73\pm 11.91$\\ 
    (IH) & $249.6\pm 26.53$ & $88.21\pm 14.11$ & \\ 
    \hline\hline 
\end{tabular}
\end{table} 

\begin{table}
\caption{The mean Bayes factors along with standard deviations 
$\langle \ln B_{\alpha\beta} \rangle\pm\sigma[\ln B_{\alpha\beta}]$
for comparing different scenarios of neutrino oscillations for
the same underlying SN neutrino emission model s20-SFHo.
Three SN distances $d=50$, 25, and 10~kpc are used and the ideal detector is assumed.
The true model $M_\alpha$ can be distinguished from the alternative $M_\beta$ at
the $>95\%$ CL for $\langle \ln B_{\alpha\beta} \rangle - 1.645\sigma[\ln B_{\alpha\beta}]>5$,
which is satisfied by all the entries except for those in bold.}
\label{tab:tab:osc_s20}
\begin{tabular}{cccc}
    \hline\hline
    $M_\alpha$\textbackslash$M_\beta$ & (NO) & (NH) & (IH)\\ 
    \hline
    $d=50$~kpc&&&\\
    (NO) &  & $\mathbf{2.66\pm 2.23}$ & $24.49\pm 7.10$\\ 
    (NH) & $\mathbf{2.95\pm 2.63}$ &  & $\mathbf{11.18\pm 4.84}$\\ 
    (IH) & $25.36\pm 7.57$ & $\mathbf{10.82\pm 4.61}$ & \\ 
    $d=25$~kpc&&&\\
    (NO) &  & $\mathbf{10.63\pm 4.46}$ & $97.96\pm 14.20$\\ 
    (NH) & $\mathbf{11.81\pm 5.25}$ &  & $44.73\pm 9.68$\\ 
    (IH) & $101.45\pm 15.14$ & $43.28\pm 9.22$ & \\ 
    $d=10$~kpc&&&\\
    (NO) &  & $66.41\pm 11.14$ & $612.26\pm 35.49$\\ 
    (NH) & $73.83\pm 13.13$ &  & $279.57\pm 24.20$\\ 
    (IH) & $634.05\pm 37.85$ & $270.52\pm 23.06$ & \\ 
    \hline\hline 
\end{tabular}
\end{table} 

\section{\label{prefactor}Analysis for Unknown SN Distance}
In Sec.~\ref{example}, assuming that the distance to the SN is known, we have 
computed the Bayesian evidence $P(D|M_\alpha)$ with a likelihood that accounts 
for both the energy-time distribution $p(E,t|M_\alpha)$ and the expected total number of 
events $\langle N \rangle_\alpha$ predicted by the SN neutrino emission model $M_\alpha$.
If the SN distance is not known, however, the expected total numbers of events 
effectively become parameters and it is practical to compare only the distributions 
$p(E,t|M_\alpha)$ and $p(E,t|M_\beta)$ to determine the distinguishability of
$M_\alpha$ and $M_\beta$. In this case, it is more difficult to distinguish the models 
because the number of observed events can no longer 
be used to provide extra discriminating power. 

For analysis of the case of unknown SN distance, we define
\begin{equation}
    \langle N \rangle = A\langle n \rangle,
\end{equation}
so that $\langle n \rangle$ contains all the
dependence on the SN neutrino emission model [see Eq.~(\ref{eq:eventratea})].
Note that $A$ only depends on the actual but
unknown SN distance and the characteristics of the assumed
detector [see Eq.~(\ref{eq:afact})].

The Bayesian evidence for $M_\beta$ is now
\begin{equation}
    P(D|M_\beta) = \prod_{i=1}^{N} p(E_i, t_i|M_\beta),
\end{equation}
where $\{E_i,t_i|i=1,2,\cdots,N\}$ denotes a signal of $M_\alpha$.
The Bayes factor for comparing $M_\alpha$ and $M_\beta$ is
\begin{equation}
    \ln B_{\alpha\beta} = \sum_{i=1}^{N} \ln \frac{p(E_i, t_i|M_\alpha)}{p(E_i, t_i|M_\beta)}.
\end{equation}
Using the Poisson distribution of $N$ with the mean $\langle N \rangle_\alpha$ and 
the energy-time distribution $\prod_{j=1}^{N} p(E_j, t_j|M_\alpha)$ for $\{E_i,t_i|i=1,2,\cdots,N\}$, 
we follow the same procedure as in Sec.~\ref{distances} to obtain
\begin{equation}
    \frac{\langle \ln B_{\alpha\beta} \rangle}{A \langle n \rangle_\alpha}= 
    \left\langle \ln \frac{p(E, t|M_\alpha)}{p(E, t|M_\beta)}\right\rangle_\alpha 
\end{equation}
and 
\begin{equation} 
    \frac{\sigma[\ln B_{\alpha\beta}]}{\sqrt{A \langle n \rangle_\alpha}}= 
    \sqrt{\left \langle \left(\ln \frac{p(E, t|M_\alpha)}{p(E, t|M_\beta)} \right)^2 \right \rangle_\alpha}.
\end{equation}

Note that $\langle \ln B_{\alpha\beta} \rangle\propto A$ and $\sigma[\ln B_{\alpha\beta}]\propto\sqrt{A}$.
We define $A_{\rm min}$ as the value of $A$ that satisfies
\begin{equation}
    \langle \ln B_{\alpha\beta} \rangle - 1.645\sigma[\ln B_{\alpha\beta}] = 5.
\end{equation}
To distinguish the true model $M_\alpha$ from the alternative $M_\beta$ at the $>95\%$ CL, we require
$A>A_{\rm min}$. The above approach follows the framework of Bayes Factor Design Analysis 
(see, e.g., \cite{Schonbrodt2018,Stefan2019}). We present the results on $A_{\rm min}$ for the assumed
ideal and SK-like detectors in Tables~\ref{tab:prefactor_ideal} and \ref{tab:prefactor_sk}, respectively. 
These $A_{\rm min}$ values all correspond to $\langle N\rangle > 50$. Monte Carlo testing suggests that 
the effects of $t_{\rm off}$ are small for $\langle N\rangle > 50$, which justifies our ignoring these 
effects in calculating $A_{\rm min}$.

For an SN at $d = 10$~kpc, $A= 2.14 \times 10^{31}$ and $1.61\times 10^{31}$ for the assumed ideal
and SK-like detectors, respectively. Either value exceeds all the corresponding $A_{\rm min}$ values in
Table~\ref{tab:prefactor_ideal} or \ref{tab:prefactor_sk}. Therefore, all of the energy-time distributions
for the adopted SN neutrino emission models can be distinguished from each other at the $>95\%$ CL with 
the signal in either detector from an SN at $d = 10$~kpc. Noting that $A\propto d^{-2}$, we can estimate
the maximum SN distance for which a pair of models can be distinguished. Take the SK-like detector
for example. 
The highest required value of $A_{\rm min}=8.26 \times 10^{30}$, necessary to distinguish the true model
z9.6-LS220(IH) from the alternative s18.6-LS220(IH), means that these two models can be distinguished for 
an SN at a distance up to $\sim 14$~kpc. 
In contrast, the lowest value of $A_{\rm min}=1.53 \times 10^{29}$, 
required to distinguish the true model s20-SFHo(NO) from the alternative z9.6-LS220(NO), means that these 
two models can be distinguished for an SN at a distance up to $\sim 103$~kpc.

\begin{table*}
\caption{Values of $A_{\rm min}$ in units of $10^{29}$ required to distinguish between the true models ($M_\alpha$) 
and the alternatives ($M_\beta$) assuming the ideal detector characteristics. 
Note that $A=2.14 \times 10^{33}({\rm kpc}/d)^2$ for the ideal detector.}
\label{tab:prefactor_ideal}
\begin{tabular}{cccccccc}
    \hline\hline
    $M_\alpha$\textbackslash$M_\beta$ & z9.6-LS220 & z9.6-SFHo & s18.6-LS220 & s18.6-SFHo & s20-SFHo & s27-LS220 & s27-SFHo\\
    \hline 
    z9.6-LS220&&&&&&&\\
\makecell{(NO)\\(NH)\\(IH)}    & \makecell{} & \makecell{11.06 \\ 9.64 \\ 7.41} & \makecell{55.92 \\ 68.68 \\ 71.60} & \makecell{11.68 \\ 10.69 \\ 8.59} & \makecell{4.22 \\ 4.40 \\ 4.01} & \makecell{7.08 \\ 9.11 \\ 11.24} & \makecell{6.04 \\ 6.07 \\ 5.12}\\ 
    z9.6-SFHo&&&&&&&\\
\makecell{(NO)\\(NH)\\(IH)}    & \makecell{10.60 \\ 9.31 \\ 7.26} & \makecell{} & \makecell{7.81 \\ 6.90 \\ 5.33} & \makecell{37.51 \\ 43.87 \\ 48.58} & \makecell{5.51 \\ 7.08 \\ 10.29} & \makecell{4.97 \\ 6.22 \\ 8.22} & \makecell{8.99 \\ 11.78 \\ 16.79}\\ 
    s18.6-LS220&&&&&&&\\
\makecell{(NO)\\(NH)\\(IH)}    & \makecell{39.66 \\ 49.08 \\ 51.62} & \makecell{5.82 \\ 5.13 \\ 3.93} & \makecell{} & \makecell{8.63 \\ 7.21 \\ 5.27} & \makecell{3.23 \\ 3.26 \\ 2.81} & \makecell{6.21 \\ 8.14 \\ 10.17} & \makecell{4.50 \\ 4.34 \\ 3.49}\\ 
    s18.6-SFHo&&&&&&&\\
\makecell{(NO)\\(NH)\\(IH)}    & \makecell{7.60 \\ 7.09 \\ 5.89} & \makecell{25.42 \\ 30.48 \\ 35.55} & \makecell{7.77 \\ 6.60 \\ 4.97} & \makecell{} & \makecell{5.27 \\ 6.99 \\ 10.79} & \makecell{5.00 \\ 6.83 \\ 11.14} & \makecell{9.44 \\ 13.42 \\ 23.01}\\ 
    s20-SFHo&&&&&&&\\
\makecell{(NO)\\(NH)\\(IH)}    & \makecell{1.45 \\ 1.62 \\ 1.64} & \makecell{1.94 \\ 2.65 \\ 4.30} & \makecell{1.55 \\ 1.65 \\ 1.57} & \makecell{2.73 \\ 3.79 \\ 6.36} & \makecell{} & \makecell{6.13 \\ 5.66 \\ 4.75} & \makecell{26.10 \\ 33.38 \\ 53.55}\\ 
    s27-LS220&&&&&&&\\
\makecell{(NO)\\(NH)\\(IH)}    & \makecell{3.19 \\ 4.38 \\ 6.08} & \makecell{2.27 \\ 3.01 \\ 4.44} & \makecell{3.89 \\ 5.43 \\ 7.57} & \makecell{3.34 \\ 4.80 \\ 8.59} & \makecell{8.15 \\ 7.52 \\ 6.29} & \makecell{} & \makecell{9.77 \\ 9.48 \\ 8.43}\\ 
    s27-SFHo&&&&&&&\\
\makecell{(NO)\\(NH)\\(IH)}    & \makecell{2.67 \\ 2.81 \\ 2.58} & \makecell{4.07 \\ 5.58 \\ 8.64} & \makecell{2.76 \\ 2.78 \\ 2.41} & \makecell{6.26 \\ 9.20 \\ 16.51} & \makecell{33.64 \\ 42.43 \\ 66.23} & \makecell{9.48 \\ 9.08 \\ 7.88} & \makecell{}\\ 
    \hline\hline 
\end{tabular}
\end{table*}

\begin{table*}
\caption{Values of $A_{\rm min}$ in units of $10^{29}$ required to distinguish between the true models ($M_\alpha$) 
and the alternatives ($M_\beta$) assuming the SK-like detector characteristics. Note that $A=1.61 \times 10^{33}({\rm kpc}/d)^2$ for the SK-like detector.}
\label{tab:prefactor_sk}
\begin{tabular}{cccccccc}
\hline\hline
$M_\alpha$\textbackslash$M_\beta$ & z9.6-LS220 & z9.6-SFHo & s18.6-LS220 & s18.6-SFHo & s20-SFHo & s27-LS220 & s27-SFHo\\
\hline
z9.6-LS220&&&&&&&\\
\makecell{(NO)\\(NH)\\(IH)}& \makecell{} & \makecell{11.40 \\ 9.91 \\ 7.61} & \makecell{60.26 \\ 75.74 \\ 82.64} & \makecell{12.06 \\ 11.03 \\ 8.86} & \makecell{4.52 \\ 4.70 \\ 4.24} & \makecell{7.73 \\ 9.98 \\ 12.34} & \makecell{6.42 \\ 6.42 \\ 5.38}\\
z9.6-SFHo&&&&&&&\\
\makecell{(NO)\\(NH)\\(IH)}& \makecell{11.01 \\ 9.64 \\ 7.51} & \makecell{} & \makecell{8.27 \\ 7.28 \\ 5.61} & \makecell{43.30 \\ 50.90 \\ 56.41} & \makecell{6.23 \\ 8.05 \\ 11.85} & \makecell{5.52 \\ 6.91 \\ 9.13} & \makecell{10.18 \\ 13.48 \\ 19.54}\\
s18.6-LS220&&&&&&&\\
\makecell{(NO)\\(NH)\\(IH)}& \makecell{42.44 \\ 53.70 \\ 59.05} & \makecell{6.08 \\ 5.34 \\ 4.09} & \makecell{} & \makecell{8.83 \\ 7.38 \\ 5.40} & \makecell{3.40 \\ 3.41 \\ 2.91} & \makecell{6.60 \\ 8.63 \\ 10.73} & \makecell{4.67 \\ 4.49 \\ 3.59}\\
s18.6-SFHo&&&&&&&\\
\makecell{(NO)\\(NH)\\(IH)}& \makecell{7.81 \\ 7.28 \\ 6.06} & \makecell{28.94 \\ 35.00 \\ 41.21} & \makecell{7.97 \\ 6.77 \\ 5.11} & \makecell{} & \makecell{5.66 \\ 7.52 \\ 11.69} & \makecell{5.31 \\ 7.26 \\ 11.89} & \makecell{10.01 \\ 14.28 \\ 24.88}\\
s20-SFHo&&&&&&&\\
\makecell{(NO)\\(NH)\\(IH)}& \makecell{1.53 \\ 1.69 \\ 1.71} & \makecell{2.13 \\ 2.93 \\ 4.85} & \makecell{1.61 \\ 1.71 \\ 1.62} & \makecell{2.88 \\ 4.02 \\ 6.82} & \makecell{} & \makecell{6.27 \\ 5.79 \\ 4.86} & \makecell{28.01 \\ 35.79 \\ 57.43}\\
s27-LS220&&&&&&&\\
\makecell{(NO)\\(NH)\\(IH)}& \makecell{3.42 \\ 4.72 \\ 6.59} & \makecell{2.45 \\ 3.26 \\ 4.83} & \makecell{4.09 \\ 5.70 \\ 7.95} & \makecell{3.50 \\ 5.04 \\ 9.09} & \makecell{8.34 \\ 7.69 \\ 6.43} & \makecell{} & \makecell{10.04 \\ 9.71 \\ 8.61}\\
s27-SFHo&&&&&&&\\
\makecell{(NO)\\(NH)\\(IH)}& \makecell{2.80 \\ 2.93 \\ 2.68} & \makecell{4.50 \\ 6.25 \\ 9.88} & \makecell{2.85 \\ 2.86 \\ 2.47} & \makecell{6.58 \\ 9.70 \\ 17.69} & \makecell{36.29 \\ 45.70 \\ 71.28} & \makecell{9.78 \\ 9.34 \\ 8.07} & \makecell{}\\
\hline\hline
\end{tabular}
\end{table*}

\section{\label{concl}Discussion and Conclusions}
Using Bayesian techniques, we have studied the feasibility of distinguishing between 
seven 1D SN neutrino emission models with the IBD events in an ideal or SK-like detector. 
For each model, the standard MSW effect with the normal or inverted neutrino mass 
hierarchy is considered along with the reference scenario of no neutrino oscillations.
We regard that the true model $M_\alpha$ can be distinguished from the alternative $M_\beta$
at the $>95\%$ CL when the mean Bayes factor and the associated standard deviation satisfy
$\langle\ln B_{\alpha\beta}\rangle-1.645\sigma[\ln B_{\alpha\beta}]>5$.
We have shown that for each of the three neutrino oscillation scenarios,
all the models can be distinguished from each other with the signal in either the ideal or 
SK-like detector from an SN at a known distance up to 25~kpc (see Tables~\ref{tab:dist25_ideal}
and \ref{tab:dist25_sk}). Some of the models could still be distinguished with an SN at a known 
distance of 50~kpc (see Tables~\ref{tab:dist50_ideal} and \ref{tab:dist50_sk}). 
We have also explored the feasibility of distinguishing between the oscillation scenarios
for a specific SN neutrino emission model. Provided that the emission model is known, for
example, from observations of the SN progenitor, these scenarios can be distinguished from
each other with the assumed ideal detector and 
an SN at a known distance of 10~kpc (see Tables~\ref{tab:osc_z9.6} and \ref{tab:tab:osc_s20}).
Finally, we have compared just the relative distributions of neutrino energy and arrival time 
predicted by the models and found that the requirement to distinguish between these distributions
can be satisfied by either the ideal or SK-like detector for an SN at an unknown distance 
up to $\sim 10$~kpc (see Tables~\ref{tab:prefactor_ideal} and \ref{tab:prefactor_sk}).

Our study covers a limited number of 1D SN neutrino emission models, but can be extended to
other 1D and multi-D models in a straightforward manner. Similarly, our study focusing on the
IBD events in water Cherenkov detectors can be generalized to other types of neutrino detectors
as well. In carrying out the present study and future ones of this kind, our goal is to estimate 
the potential of current and planned neutrino detectors to distinguish between various SN models.
In the event of an actual SN, similar Bayesian techniques to those presented here can be used to
rank various SN neutrino emission models as discussed in \cite{Olsen2021} for the case of SN~1987A.
In addition, a $p$-value test can be performed to check if a model is incompatible with the data
\cite{Olsen2021}. Based on the results presented here, it is very 
likely that the neutrino 
signal from the next Galactic SN would allow us to differentiate a wide range of models.

\begin{acknowledgments}

We thank the Garching group for giving access to their SN neutrino emission models. J.O. is grateful to Ermal Rrapaj and Andre Sieverding for useful discussions. This work was supported in part by the US Department of Energy under grant DE-FG02-87ER40328. Calculations were carried out with resources of the Minnesota Supercomputing Institute.

\end{acknowledgments}

\end{document}